\documentclass[prd,
twocolumn,superscriptaddress,preprintnumbers,nofootinbib]{revtex4}
\usepackage{graphicx}
\usepackage{epsfig}
\usepackage{bm}
\usepackage{latexsym,amssymb,amsmath,amssymb,wasysym,float}

\usepackage{color}

\usepackage{scrextend}
\usepackage{mathrsfs}
\usepackage{enumitem}

\usepackage[usenames,dvipsnames]{xcolor}
\definecolor{orange}{cmyk}{0,0.5,1,0}
\definecolor{rossoCP3}{cmyk}{0,.88,.77,.40}
\definecolor{graa}{rgb}{0.8,0.8,0.8}
\definecolor{blaa}{rgb}{0.2,0.2,0.6}

\definecolor{darkgreen}{rgb}{0.,0.6,0.}

\begin{document}

\preprint{MPP-2025-203}
\preprint{LMU-ASC 24/25}

\vskip1cm

\title{
{~}\\{~}\\
Species Quantum Mechanics\\{~}\\
}

\author{\bf Luis A. Anchordoqui}

\affiliation{Department of Physics and Astronomy,  Lehman College, City University of
  New York, NY 10468, USA
}

\affiliation{Department of Physics,
 Graduate Center,  City University of
  New York,  NY 10016, USA
}

\affiliation{Department of Astrophysics,
 American Museum of Natural History, NY
 10024, USA
}

\author{\bf Dieter\nolinebreak~L\"ust}

\affiliation{Max--Planck--Institut f\"ur Physik,  
 Werner--Heisenberg--Institut, 85748 Garching, Germany}

\affiliation{Arnold Sommerfeld Center for Theoretical Physics, 
Ludwig-Maximilians-Universit\"at M\"unchen,
80333 M\"unchen, Germany
}

\author{\bf Severin L\"ust}
\affiliation{Sorbonne Université, CNRS, Laboratoire de Physique Théorique et Hautes Énergies, LPTHE, F-75005 Paris, France
}

\begin{abstract}
  \noindent {~}\\
  In this note we take the initial steps towards a formulation of {\sl Species Quantum Mechanics}. Specifically, we consider quantum operators that correspond to
the species number $N_s$ and the tower mass scale $m_t$ in the context of the swampland distance conjecture. We discuss the commutation relations, a possible wave function, and
symplectic duality transformations on the conjugate variables. Furthermore, 
we argue that the Castellano-Ruiz-Valenzuela (CRV) pattern is a consequence of the
canonical commutation rules of moduli space quantum mechanics. 
We also connect the canonical quantization to the periods
of ${\cal N}=2$ Calabi-Yau compactifications to explore other
aspects of the CRV pattern, including 
its possible connection to the Ooguri-Vafa-Verlinde black hole
quantization procedure. 

\end{abstract}

\date{October, 2025}
\maketitle

\section{Introduction -- General Idea}

In quantum mechanics non-commuting operators correspond to conjugate
variables. The most famous example is the pair of position $\hat x$ and
momentum $\hat p$ operators, which are non-commutative and form the basis of
Heisenberg's uncertainty principle~\cite{Heisenberg:1927zz}. In its
modern form, the Robertson uncertainty relation for two non-commuting observables $ A$ and $ B$ is defined by
\begin{equation}
  \Delta A \cdot \Delta  B \geq \frac{1}{2} |\langle [\hat A,\hat B]\rangle | \,,
\label{UP}
\end{equation}
where $(\Delta A)^2 = \langle  A^2 \rangle - \langle   A \rangle^2$
denotes the variance of the operator $\hat A$ in a given quantum state $\rho$, and
$\langle \cdot \rangle = {\rm Tr}\{\rho \cdot\}$ indicates the expectation value~\cite{Robertson:1929zz}. The commutator $[\hat A,\hat B]$ captures the fundamental incompatibility of the
operators representing observables $A$ and $B$. Then for canonical
conjugate pairs (e.g., $\hat x$ and $\hat p$) their non-zero
commutator $([\hat x,\hat p] = i \hbar)$ guarantees that a fundamental minimum uncertainty exists.

Applying the quantum rules to gravity is still a big challenge in theoretical physics. Normally, one expects that effects from quantum gravity become important at the Planck scale $M_p\simeq 10^{19}{\rm GeV}$. However, it has been conjectured that quantum gravity effects could manifest
at scales much lower than the Planck scale~\cite{Arkani-Hamed:1998jmv,Antoniadis:1998ig,Randall:1999ee,Montero:2022prj}. In fact, in the last two decades this conjecture has seen an intense Swamplandish
revival~\cite{Vafa:2005ui}. The essence of this resurgence
can be encapsulated as follows. The principle of naturalness has
historically guided progress in High Energy
Physics. Indeed, the effective field theory (EFT) approach provides a powerful framework for describing physical phenomena at a
particular energy scale. A key characteristic of an EFT is its finite
domain of validity, defined by a cutoff energy. The concept of naturalness implicitly assumes that
heavy, integrated-out fields do not contribute significantly to
low-energy physics. While naturalness arguments have proven over time useful in
High Energy Physics, they are now being challenged by the significant
fine-tuning problems posed by the Higgs  and the cosmological constant. In this vein, recent developments in
theoretical physics advocate for replacing the naturalness logic with
guiding principles derived from the demand for ultra-violet (UV) consistency of EFTs. This is
particularly motivated by the requirement that such EFTs must be
consistent with a valid UV completion in a theory of quantum gravity, an essential consideration due to the presence of gravitational
interactions in our universe. The Swampland conjectures, which outline
the conditions for an EFT to be compatible with quantum gravity, provide concrete examples of how UV consistency dictates the parameters and structures of low-energy physics~\cite{Palti:2019pca,vanBeest:2021lhn,Agmon:2022thq}.

In particular, the swampland {\it distance conjecture} (DC)~\cite{Ooguri:2006in} together with the {\it emergent string conjecture} (ESC)~\cite{Lee:2019wij}
provide a 
concrete mechanism by which an EFT drastically breaks down. The DC predicts the emergence of an infinite tower
of states, with their masses decreasing exponentially as the geodesic
distance traveled in the field space grows. In addition, the ESC is
stating that the towers of species that become light in some infinite
distance limit can be either light Kaluza-Klein (KK) modes, 
corresponding to a decompactification limit, or light string states, belonging to a weakly coupled string limit. The large number of states in the tower implies the breakdown of semiclassical Einstein
gravity at a cut-off
scale, the quantum gravity scale commonly also dubbed the species
scale
$\Lambda_s$~\cite{Dvali:2007hz,Dvali:2007wp,Dvali:2009ks}.\footnote{Other
  aspects of species were discussed
  in~\cite{Dvali:2019jjw,Dvali:2019ulr,Dvali:2020wqi,Dvali:2010vm,Dvali:2008fd,Dvali:2012uq,Dvali:2021bsy}.}
This scale can be expressed in terms on the number $N_s$ of species
below $\Lambda_s$ in the following way in Planck units:
\begin{equation}
\Lambda_s=N_s^{1\over 2-d}\, . \label{species}
\end{equation}
Another important aspect of the distance conjecture is the existence
of duality symmetries, which connect different infinite distance
limits to each other. In fact, the emergence of duality symmetries together with finiteness and compactifiability of the moduli space 
can be seen from certain bottom-up
arguments~\cite{Delgado:2024skw}. Because of the emergence of duality
symmetries the DC is also sometimes called the {\it duality
  conjecture}. Further aspects of the DC in relation to information theory were discussed in~\cite{Stout:2021ubb,Stout:2022phm}.
Finally, a complementary discussion of the DC in terms of flow equations can be found in~\cite{Kehagias:2019akr}.

Another generic and important feature of quantum gravity and string theory is that the moduli that describe families of quantum gravity vacua are given in terms of vacuum expectation values of scalar fields $\phi$.
Actually, in the EFT these scalar fields $\phi$ should be treated as quantum fields, denoted by $\hat\phi$. 
Together with the conjugate momentum operator $\hat \pi$ they act on the quantum mechanical Hilbert space.
Furthermore, the fields $\hat\phi$ and $\hat\pi$ satisfy some canonical commutation relations of the form
\begin{equation}
\lbrack \hat\phi,\hat\pi\rbrack =i\hbar \, .
\end{equation}

Looking at the EFT of quantum gravity, it turns out that most of the
couplings, operators, and quantities therein, in general denoted by
${\cal O}$, are field dependent, i.e. ${\cal O}={\cal O}(\phi)$. Therefore, they also should be treated as quantum mechanical operators $\hat{\cal O}(\hat \phi)$.
 More generally, operators that depend both on $\hat\phi$ and on $\hat\pi$ should obey some quantum mechanical commutation relations with commutators of the form
$\lbrack\hat{\cal O}_1,\hat{\cal O}_2\rbrack\simeq\hat{\cal O}_3$.  
For example, the KK mass scale, or more generally the
mass scale $m_t$ appearing in the DC, are scalar field dependent, as
they are functions of the scalars that correspond to geometric moduli and the string coupling constant,
which we will collectively denote by $t_i$. The same holds also for
the field dependent species number ${N}_s(t)$
\cite{vandeHeisteeg:2022btw}. Hence, the tower mass scale ${m}_t$ as well as $N_s$ correspond to some quantum operators
$\hat m_t(\hat t)$ and $\hat N_s(\hat t)$.

{\sl Species Thermodynamics}  states that species enjoy thermodynamic properties~\cite{Cribiori:2023ffn,Basile:2023blg,Basile:2024dqq,Herraez:2024kux,Herraez:2025clp}.
The species entropy ${\cal S}_s$ is determined by an area law, it is given in terms of the number of species in a given tower, i.e. ${\cal S}_s=N_s$,  and it can be derived from the entropy of corresponding minimal black holes:
${\cal S}_s={\cal S}_{\rm BH}$~\cite{Cribiori:2022nke}. 
Species thermodynamics also provides bottom-up arguments for the
emergence distance conjecture. Furthermore, species thermodynamics
provides information about the classical dynamics of $N_s$ and $m_t$,
as for example, the second law of species thermodynamics states that the number of species must not decrease, i.e. $\delta N_s\geq0$.

In this note we take the next steps after the introduction of species thermodynamics to lay a foundation of {\sl Species
  Quantum Mechanics}. We will discuss some aspects of the
quantum properties of $\hat m_t$ and $\hat N_s$, together with
some related commutation and uncertainty relations. 
In this way, species quantum mechanics will constitute a kind of mini-superspace version of quantum gravity by considering the quantum Hilbert space of the species.

Recently, Castellano, Ruiz, and Valenzuela (CRV) pointed out that in
infinite-distance limits there is a universal pattern linking the
(logarithmic) 
gradients of the mass gap $m_t$ of the lightest tower to the species
scale $\Lambda_s$,
\begin{equation}
\frac{\nabla m_t}{m_t} \cdot \frac{\nabla \Lambda_s}{\Lambda_s} =
\frac{1}{d-2} \,,
\label{CRV}
\end{equation}
where $d$ is the number of dimensions of the lower dimensional
EFT~\cite{Castellano:2023stg,Castellano:2023jjt}. In terms of the number of species, using eq.(\ref{species}) the CRV
pattern can be recast as
\begin{equation}
\frac{\nabla m_t}{m_t} \cdot \frac{\nabla N_s}{N_s} = -1 \,.\label{CRVpattern}
\end{equation}
Note that this relation is independent from the number of space-time dimensions.

In our paper we first  identify the pair of conjugate variables of the EFT with
 moduli space ${\cal M}_\phi$.
Then, we
 will show that
$N_s$ and $m_t$ are themselves closely related to conjugate variables and we will discuss the commutation relations involving these operators.
This will allow us  to relate the  pattern (\ref{CRV}) or (\ref{CRVpattern}) to 
a particular commutation relation of species quantum mechanics. 
We will discuss the commutation relations involving the operators $N_s$ and $m_t$, a possible wave function, and
symplectic duality transformations on the conjugate variables. 
Finally,
using
the periods of ${\cal N}=2$
Calabi-Yau (CY) compactifications  we make contact with the quantization
rules set up by Ooguri, Vafa, and
Verlinde (OVV) in~\cite{Ooguri:2005vr}. 
A synopsis   of
our arguments as well as some important caveats
are discussed at the end of the summary section.

\section{Species quantum mechanics}

Let us consider a $d$-dimensional EFT of the (schematic) form:
\begin{eqnarray}
S_{\rm EFT}&\sim&\int d^dx \sqrt{-h}\biggl(M_p^{d-2}\bigl( R + {1\over 2}g_{ij}(t){\partial_\mu t^i\partial_\mu t^j}+V(t)\nonumber\\
&+&\sum_{n\geq 1}c_nN_s^{n-2\over d-2}(t){\cal O}(R^{2n})\bigr)\nonumber\\&+&\sum_{n>d}\tilde c_nm_t^{d-n}(t)\tilde{\cal O}(R^{2n})+\dots\biggr)
 \,,
\end{eqnarray}
where $h$ refers to the space-time metric, the scalar fields $t^i$ correspond to the moduli of the string compactification, $g_{ij}(t)$ is the moduli space metric and $V(t)$ a possible moduli dependent scalar potential. In addition,
this action 
is a double EFT expansion \cite{Calderon-Infante:2025ldq} with respect to the two scales that are relevant for our discussion: there is a first set of higher derivate operators ${\cal O}(R^{2n})$ that are determined by the 
moduli dependent species number $N_s(t)$~\cite{vandeHeisteeg:2022btw}. On the other hand, the second set of higher derivative operators $\tilde {\cal O}(R^{2n})$
is suppressed by the tower mass scale $m_t(t)$.

\subsection{Canonical variables and commutation relations}

We will now treat the scalar fields $t^i$ as quantum fields, and we first want to determine their conjugate momenta $\pi_t$.
Following the standard canonical quantization procedure, the conjugate
momenta in quantum field theory are given by 
$
\pi_{i,\mu}={\delta {\cal L}_{\rm EFT}\over \delta(\partial_{\mu}
  t_i)}=g_{ij} \ \partial_{\mu}t^j$.
But to obtain a sensible quantization, i.e. in order to describe a
quantum mechanical system, the canonical momenta should only be defined w.r.t to the time derivative. Therefore, as also considered in the context of finiteness \cite{Delgado:2024skw}, 
 we assume that  all spatial coordinates are compactified and only the
 time coordinate $\tau$ is left.\footnote{Actually, in a way we do not
   really compactified to one dimension, but we just ignore the three
   space-like dimensions. A subtle point of this truncation is that we
   work in four-dimensional Planck units and so the relations between
   the tower masses and the species scale derived in the
   four-dimensional EFT do not change.} Then 
following the standard canonical quantization procedure, 
the canonical momenta are just given by the time-derivative of the moduli fields:
\begin{equation}
\pi_{t,i}={\delta {\cal L}_{\rm EFT}\over \delta(\partial_{\tau} t_i)}
=g_{ij} \ \partial_\tau t^j=g_{ij} \ \dot t^j\, .\label{momentum}
\end{equation}
This leads to an Hamiltonian of the following form:
\begin{equation}
{\cal H}={1\over 2}g^{ij}(t)\pi_{t,i}\pi_{t,j}+V(t)\, .\label{hamilton}
\end{equation}

The canonical quantization of the quantum mechanical moduli space is now pursued by imposing the following canonical equal-time commutation relation (we have set $\hbar=1$ in all the following commutators)
on the pair $(t^i,\pi_{t,j})$ of conjugate variables:\footnote{We could also consider non-equal-time commutators of the form
$\lbrack t^i(\tau),t^j(\tau')\rbrack=i G^{ij}(\tau,\tau')$ and $\lbrack
t^i(\tau),\pi_{j}(\tau')\rbrack=i g_{jk} \partial_{\tau'} G^{ik}(\tau,\tau')$ for a suitable commutator function $G^{ij}$.}
\begin{equation}
\lbrack t^i,\pi_{t,j}\rbrack =\lbrack t^i,g_{jk}\dot t^k\rbrack =i  \delta^i_j\, .
\end{equation}

Let us now proceed by computing the commutation relations between some
functions of the operators $t^i$ and $\pi_{t,i}$. 
Specifically, we are
interested in the species number operator
$N_s$, in the tower scale operator $m_t$ and 
in the commutation relations that involve these operators.
Bearing this in mind, we relate $N_s$ and $m_t$ to the canonical variables $t^i$ and $\pi_j$ in a concrete way. To do that we assume that we are close to the boundary of the moduli space, where the infinite distance conjecture holds~\cite{Ooguri:2006in}.
This conjecture relates the tower mass scale $m_t$ to the moduli space metric $g_{ij}$ in the following way:
\begin{equation}
m_t(t)=e^{-\alpha d(t)}\, ,
\end{equation}
where $d(t)$ is the geodesic distance from an arbitrary reference point $t_0$ to the point labeled by $t$, for asymptotically
large $t$.
It is computed from the moduli space metric as (here we assume that $V(t)=0$)
\begin{equation}d(t)=\int_0^1 d\lambda \ \sqrt{{dt^i\over d\lambda} \
    g_{ij}(\lambda) \ {dt^j\over d\lambda}}\ ,
\end{equation}
where $\lambda$ is a parameter of the path, with $\lambda =0$
corresponding to $t_0$ and $\lambda = 1$ to $t$. For simplicity, we now assume that we are dealing with a tower of species
with a tower mass scale 
\begin{equation}
m_t=1/t^\alpha\, .\label{mtower}
\end{equation}
For the case that the species are given in terms of KK modes corresponding to $p$ compact dimensions, the parameter $\alpha$ is given as~\cite{Castellano:2021mmx}
\begin{equation}
\alpha={p+2\over 2p}\, 
\end{equation}
The corresponding metric $g_{tt}$ is given by 
\begin{equation}
g_{tt}=\alpha/t^2\, .\label{metric}
\end{equation}
The case $p=\infty$  corresponds to the case of a tower of light string excitations. In the next section on Calabi-Yau periods we will discuss three cases in more detail:
 the emergent string limit with $\alpha=1/2$, the decompactification to six dimensions with $\alpha=1$, and the decompactification to five dimensions with $\alpha=3/2$.
 
 Using the metric (\ref{metric}) we can easily determine the canonically normalized scalar field $\phi$:
 \begin{equation}
 \phi=\sqrt\alpha\log t\, .
\label{canonicallyphi}
\end{equation}
 The canonical momentum is then simply given by
 \begin{equation}
 \pi_\phi=\dot\phi
 \end{equation}
 and satisfies the commutation relation
 \begin{equation}
 \lbrack
 \phi,\dot\phi\rbrack=i\, .
 \end{equation}

Next, as we discuss below, for KK compactifications, the number of species is proportional to
the volume of the compact space in string units, and thus just given by $t$ itself, i.e. 
\begin{equation}
N_s= t \, .
\label{speciesnumber}
\end{equation}
Thus, together with (\ref{metric}), the canonical momentum $\pi_{N_s}$, conjugate to $N_s$, can be written as
\begin{equation}
\pi_{N_s}={\alpha\over N_s^2}\dot N_s\, .
\end{equation}
This is of course consistent with the time evolution of $N_s$ in quantum mechanics, which is given by the following commutator:
\begin{equation}
\dot N_s=i\lbrack {\cal H},N_s\rbrack =g_{tt}^{-1}\pi_{N_s}\, .
\end{equation}

Next, we like to express $\pi_{N_s}$ by the tower mass scale $m_t$ and after a short computation we obtain
\begin{equation}
\tilde m_t:=\pi_{N_s}=-m_t^{1-\alpha\over\alpha}\dot m_t\, .
\end{equation}
Note that for $\alpha=1$, $\tilde m_t=-\dot m_t$.
Hence, assuming that the swampland distance conjecture together with
the 
given form of the species towers holds, the pair of operators $(N_s,\tilde m_t)$  are canonically conjugate to each other:
\begin{eqnarray}
\lbrack N_s(t), \tilde m_t(t)\rbrack=-
m_t^{1-\alpha\over\alpha}\lbrack N_s(t), \dot m_t\rbrack=
i\, .\label{fhcom10}\
\end{eqnarray}
We see that for $\alpha=1$ the pair $(N_s,-\dot m_t)$ are conjugate to each other:
\begin{eqnarray}
 \lbrack N_s, -\dot m_t\rbrack=
i\, .\label{fhcom11}\
\end{eqnarray}

The commutator (\ref{fhcom10}) leads to a Robertson uncertainty relation of the form
\begin{equation}
 \Delta N_s  ~\Delta\tilde m_t\geq {1\over 2}\, .
\end{equation}
The interpretation of this Robertson uncertainty is very interesting:
it means that if one does a precise measurement of the number of species, or respectively the species scale, quantum mechanics forbids to measure the time variation of the tower mass scale with high precision. Of course the reverse statement also is true.

\subsection{The CRV pattern}

Let us now investigate the CRV pattern in more detail.
First we note that
 commutator between the two functions $f(t)$ and $\dot h(t)$
 can be evaluated as follows:
\begin{eqnarray}
\lbrack f(t),{d\over d\tau} h(t)\rbrack&=& \lbrack f(t),\partial_ih(t)\dot t^i\rbrack \nonumber\\
&=&\lbrack f(t),\partial_i h(t)g^{ij}(t)\pi_j\rbrack\nonumber\\
&=&\partial_i h(t)g^{ij}(t)\lbrack f(t),\pi_j\rbrack\nonumber\\
&=& 
\partial_i f(t)g^{ij}(t) \partial_k h(t)\lbrack t^k,\pi_j\rbrack\nonumber\\
&=& i~
\partial_i f(t)g^{ij}(t) \partial_j h(t)\, .\label{fhcom5}
\end{eqnarray}
So in short we get
\begin{eqnarray}
\lbrack f(t),\dot h(t)\rbrack=i~
 \nabla f(t)\cdot \nabla h(t)\, .\label{fhcom}
\end{eqnarray}
It corresponds to a Robertson uncertainty relation of the following form
\begin{eqnarray}
\Delta f(t)~\Delta\dot h(t)\geq
\frac{1}{2} |\langle\nabla f(t)\cdot \nabla h(t)\rangle|\, .\label{fhuncer}
\end{eqnarray}

Now, let us apply this  to the functions, which correspond to the tower mass scale and number of species in the CRV pattern. Specifically, we set $f(t)=\log N_{s}(t)$ and $h(t)=\log m_t(t)$.
Then the commutator (\ref{fhcom}) becomes:
\begin{eqnarray}
\lbrack \log N_s(t), {d\over d\tau}\log m_t(t)\rbrack=i~
 {\nabla N_s(t)\over N_s(t)}\cdot {\nabla m_t(t)\over m_t(t)}\, .\label{fhcom1}
\end{eqnarray}
We see that the r.h.s. of this equation is just the expression as given by the CRV pattern. So assuming that the CRV pattern holds we can conclude that the commutator between these two operators is constant:
\begin{eqnarray}
\lbrack \log N_s(t), {d\over d\tau}\log m_t(t)\rbrack=
 -i\, .\label{fhcom2}
\end{eqnarray}
In this case $N_s$ and the time derivative of $m_t$ are related to canonically conjugate operators.

One can also reverse the argument by imposing that $\log N_s(t)$ and ${d\over d\tau}\log m_t(t)$ are conjugate operators.  Then 
the CRV pattern, written as an operator product,  is a consequence of the canonical
commutation rules of moduli space quantum mechanics.
Let us demonstrate the equivalence  relation between the CRV pattern and the considered pair of conjugate operators in an explicit way.
For that we use the relations between $N_s$ and $m_t$ to the canonical variables $t^i$ and $\pi_j$,
which we have introduced before.
Then, we can directly evaluate the  l.h.s. of \eqref{fhcom1}:
\begin{eqnarray}
\lbrack \log N_s(t), {d\over d\tau}\log m_t(t)\rbrack&=&{1\over t}\lbrack t,{d\over d\tau}\log m_t(t)\rbrack \nonumber\\
&=&{1\over t}\lbrack t,{\partial_t m_t(t)\over m_t(t)}\dot t\rbrack\nonumber\\
&=& -{\alpha\over t^2}\lbrack t,g_{tt}^{-1} \pi_t \rbrack\nonumber\\
&=&-\lbrack t,\pi_t \rbrack=-i\, .\label{fhcom7}\
\end{eqnarray}
So we see that the CRV pattern (\ref{CRVpattern}) is indeed a consequence of the basic commutation rule.

The CRV pattern looks particularly simple in terms of $\phi$ and $\dot\phi$.  Specifically, one has that
\begin{equation}
\log N_s(t)={\phi\over\sqrt{\alpha}}\, ,\quad {d\over d\tau}\log m_t=-\sqrt{\alpha}\dot\phi\, .
\end{equation}
So in the canonical field basis $\log N_s$ and $\log m_t$ are directly related to the conjugate variables and the CRV pattern can be directly expressed as
\begin{eqnarray}
\lbrack \log N_s(t), {d\over d\tau}\log m_t(t)\rbrack=-\lbrack \phi,\dot\phi \rbrack=-i\, .\label{fhcom8}\
\end{eqnarray}

Finally, we see that ${d\over d\tau}\log m_t$ and $\log N_s$ obey a quantum mechanical  uncertainty relation of the form 
\begin{equation}
 \Delta \biggl(\log N_s \biggr) \cdot \Delta\biggl({d\over d\tau}\log m_t\biggr )\geq {1\over 2}\, .
\end{equation}

All the above relations were shown to hold at the asymptotic boundaries of the moduli space.
Inside the moduli space the moduli dependence of $N_s$, $m_t$ as well as the 
CRV pattern is expected to receive corrections. For example, the tower
mass scale $m_t$ may be replaced by the so-called black hole scale
$\Lambda_{\rm BH}$~\cite{Bedroya:2024uva}, where certain black hole phase transitions take place.
In addition, some explicit corrections were computed in~\cite{vandeHeisteeg:2022btw,vandeHeisteeg:2023dlw,Basile:2025bql}.
Going   to string compactifications with a higher number of
unbroken supersymmetry, these corrections can be computed or are under
better control. For example, this is the case in nine or in ten dimensional type string vacua, or also for ${\cal N}=2$ string compactifications, which we will consider in the next section.
An alternative method to obtain some information about the interior of the moduli space from its boundary was discussed in~\cite{Lust:2024aeg}.

\subsection{Symplectic duality transformations}

Within the swampland approach, duality transformations typically
relate different, but often physically equivalent infinite distance
limits to each other. Swampland arguments, as well as the related
finiteness and compactifiabilty of moduli space,
how duality symmetries lead to bottom-up predictions of duality
symmetries, were recently given in \cite{Delgado:2024skw}. Examples are the well-known T-duality transformations
of the form $t\rightarrow 1/t$, which exchange the KK tower with the
stringy winding tower, or the S-duality that relates weak and strong coupling to each other.

On the other hand, in quantum mechanics symplectic transformations on a pair of conjugate variable $(q,\pi)$ of the form
\begin{equation}
q\rightarrow \pi\, ,\quad \pi\rightarrow -q
\end{equation}
are linear canonical transformations that preserve the commutation relations and act as unitary transformations on the Hilbert space.

Now, we want to argue that the duality transformations in the string moduli space have their origin in the symplectic transformation
of the canonical variables of the species quantum mechanics.
In particular T-duality should correspond to a symplectic transformation of the form
\begin{equation}
t\rightarrow \pi_t\, ,\quad \pi_t\rightarrow -t \,,
\end{equation}
with $\pi_t={\alpha\over t^2}\dot t$,
or respectively to
\begin{equation}
N_s\rightarrow \pi_{N_s}=\tilde m_t\, ,\quad \pi_{N_s}=\tilde m_t\rightarrow - N_s\label{symplectic}
\end{equation}
with $\tilde m_t=-m_t^{1-\alpha\over\alpha}\dot m_t$.
This can be best seen for the case $\alpha=1$, where $N_s=t$ and $m_t=1/t$, and $\lbrack N_s,\dot m_t\rbrack=-i$.  Here, the T-duality transformation $t\rightarrow 1/t$  directly exchanges $N_s$ and $m_t$. Hence, it is equivalent
to the symplectic transformation in \eqref{symplectic}.
Actually, these are the cases in which the symplectic transformation acts as a symmetry transformation on the Hilbert space, and T-duality corresponds to a self-duality on the string spectrum. An analogous behaviour is for example expected for
the $\mathrm{SL}(2,\mathbb{Z})$ duality transformation of type IIB string theory.

We will also mention the connection between T-duality and symplectic transformations on the Calabi-Yau periods in the next section.

\subsection{Species wave function}

So far we only discussed some operators and their commutation relations of species quantum mechanics, but not the Hilbert space, respectively the quantum mechanical wave functions, on which the operators act.
This is of course of importance for the dynamics of the tower of species, i.e. how the number of species and the tower mass scale evolves in time. This should be  compared with the 
classical dynamics of the species, as e.g. described by some flow equations or also arguments coming from species  thermodynamics.

It has long been known that the relevant field equation in
  quantum gravity is the Wheeler-DeWitt equation
  (WDW)~\cite{Wheeler:1968iap,DeWitt:1967yk}. More concretely, in the
  Arnowitt–Deser–Misner (a.k.a. ADM) formalism~\cite{Arnowitt:1959ah} the WDW equation takes the form
\begin{equation}
  {\cal H} \Psi \lbrack h \rbrack =0 \,,
\end{equation}
where $h$ is the space-time metric and the Hamiltonian is given by
 \begin{equation}
  {\cal H}= {1\over\sqrt h}\left( \pi^{ij} \pi_{ij} -{1\over 2}\pi^2
   \right)-\sqrt {h} \ R \, .
\end{equation}
Note that time and time derivatives do not appear in the WDW equation, i.e. no time evolution emerges in terms of the WDW wave function dependence.\footnote{Some ideas about the wave function of the universe in brane world models are given in~\cite{Anchordoqui:2000du}.} 

However, herein we are not discussing the quantum evolution of the $d$-dimensional space-time metric, but only the part from the internal geometry in the form of the matter moduli fields  $\phi$.
This amounts to replace ${\cal H}$ by ${\cal H}_h + {\cal H}_\phi$ and the wave function
will also depend of $\phi$, {\it viz.},  $\Psi\lbrack h,\phi\rbrack$. Moreover, in the presence of the extra field $\phi$, one need to define time and a corresponding energy for $\phi$.
Then, in the WKB approximation, the WDW equation turns into an
Hamilton-Jacobi equation for $\phi$, see
e.g.~\cite{Halliwell:1989myn}. This Hamilton-Jacobi equation can be regarded as coming from a
corresponding Schr\"odinger equation. Therefore, we assume that the
dynamics of the matter is determined by the wave function $\psi_s(t)$
of species quantum mechanics, which should correspond to some ``wave
function of the universe'' satisfying a 
Schr\"odinger equation of the form
\begin{equation}
{\cal H} \psi_s=E \psi_s\, ,\label{schrodinger}
\end{equation}
with the Hamiltonian as given in (\ref{hamilton}).
The solutions of the Schr\"odinger equation will provide the expectations values for observables $\hat{\cal O}(t)$:
\begin{equation}
\langle \hat{\cal O}(t)\rangle=\langle \psi_s|   \hat{\cal O}(t)  |\psi_s\rangle\,.
\end{equation}
For example the  expectation values of the number of species, $\langle N_s\rangle =\langle \psi_s|N_s|\psi_s\rangle$, or of the species scale, 
$\langle \Lambda_s\rangle =\langle \psi_s|\Lambda_s|\psi_s\rangle$,
are important observables that can be measured in principle  by experiments.

The kinetic term of this Hamiltonian is just the Laplace operator on
the moduli space ${\cal M}$.\footnote{Eigenfunctions of the Laplacian
  operator were discussed in~\cite{Aoufia:2025ppe}.} It can be most conveniently expressed in terms of the canonical field basis:
\begin{equation}
{\cal H}={1\over 2}\pi_\phi\pi_\phi\,.
\end{equation}
Using that the operator $\pi_\phi$ can be expressed as $\pi_\phi=-i{\partial\over \partial\phi}$
the Hamiltonian becomes
\begin{equation}
{\cal H}=-{1\over 2}{\partial^2\over \partial\phi^2}\,.
\end{equation}

For the  case of a vanishing potential, $V(t)=0$, the wave function has a similar form as the one of a free particle, namely 
\begin{equation}
\psi_s(t)\sim \exp(i k_\phi \phi)\, ,
\label{psit1}
\end{equation}
where $k_\phi$ is the momentum eigenvalue of $\pi_\phi$. With $\phi=\sqrt\alpha\log t=\sqrt\alpha\log N_s$, the species wave function is proportional to 
\begin{equation}
\psi_s(t) \sim \exp\left[i\sqrt\alpha k_\phi\log N_s(t) \right]\,. 
\end{equation}
This expression provides an interesting connection  to the species thermodynamics: since $N_s$ corresponds to the species entropy, namely
to the entropy of a minimal black hole, i.e. $N_s={\cal S}_{\rm BH}$, we can write the wave function as
\begin{equation}
\psi_s(t)\sim \exp(i \sqrt\alpha k_\phi\log{\cal S}_{\rm BH})\, .
\end{equation}
For supersymmetric black holes, which are following the attractor mechanism, the black hole entropy as well as the scalar fields at the horizon are fully determined by a set of electric and magnetic charges $(Q,P)$:
$t=t(Q,P)$ and ${\cal S}_{\rm BH}={\cal S}_{\rm BH}(Q,P)$.
So the species wave function will eventually be proportional to an expression of the form
\begin{equation}
\psi_s(P,Q)\sim \exp\bigl[i\sqrt\alpha k_\phi \log{\cal S}_{\rm BH}(Q,P)\bigr]\, .\label{specieswave}
\end{equation}
This wave function resembles some similarity with the OVV wave function of the universe, which we will mention in the next section.
Note also that the charges $(P,Q)$ undergo symplectic transformations, like $N_s$ and $\pi_t$ do.

Before proceeding, a brief word of caution is necessary.  While the plane wave given in (\ref{psit1}) is a definite-momentum solution to the Schr\"odinger equation, it is not a localized solution. Actually, $\psi_s$ as given in (\ref{psit1}) is not normalizable over infinite moduli space. Thus, to determine e.g. $\langle N_s \rangle$ the wave function $\psi_s$ must be a wave packet with a finite extent in moduli space (e.g. a Gaussian wave packet), which is a superposition of different momentum (or wave number $k_\phi$) eigenstates,
\begin{eqnarray}
\psi_s(t) & \sim & \frac{1}{\sqrt{2\pi}} \int_{-\infty}^{+\infty}
                   \varphi (k_\phi) \ e^{ik_\phi \phi(t)} \ dk_\phi \nonumber \\
& \sim & e^{-\frac{[\phi(t)-\phi_0]^2}{2 \sigma^2}} \ e^{ik_{\phi_0} \phi(t)} \,,
\end{eqnarray}
where $\sigma$ determines the width of the packet and the expectation
value  $\langle \phi \rangle \sim \phi_0$ aligns with the classical
trajectory. The expectation value then evolves over time according to Ehrenfest's theorem, where its velocity is related to the average momentum $\langle \pi_\phi \rangle = k_{\phi_0}$. Using (\ref{canonicallyphi}) and (\ref{speciesnumber}) it is now straightforward to obtain $\langle N_s \rangle$.

We can also consider the case with a non-vanishing potential $V(t)$.  
At the classical level, the distance conjecture and the classical trajectories were recently investigated 
in several works~\cite{Lust:2019zwm,Kehagias:2019akr,Li:2023gtt,Basile:2023rvm,Palti:2024voy,Demulder:2023vlo,Mohseni:2024njl,Debusschere:2024rmi,Demulder:2024glx,Grimm:2025cpq}.
E.g. a refined  distance  measure \cite{Mohseni:2024njl}
\begin{equation}d(t)=\int d \lambda \sqrt{{dt^i\over d \lambda }g_{ij}(\lambda ){dt^j\over d \lambda}+2V(t(\lambda))}\, .
\end{equation}
can be used to determine the geodesics distance in parameter space in the presence of a potential.
It would be interesting to compare the classical trajectories $t_{\rm
  cl}(\tau)$ with the quantum trajectories $t_{\rm qm}(\tau)$, which we will discuss in the following.

For that let us assume that the potential follows the anti-de Sitter
conjecture (ADC)~\cite{Lust:2019zwm}, which states that the value of the potential at its minimum is proportional to some power of the tower
mass scale $m_t$. Here let us assume the entire function $V(t)$ follows the ADC, which leads to
\begin{equation}
V(t)=c~\left[m_t(t) \right]^n\, ,
\end{equation}
where $n$ is a positive parameter of order one, and $c$ is negative for anti-de Sitter spaces and positive for de Sitter spaces. E.g. in many non-scale separated supersymmetric AdS vacua $n=2$. 
With (\ref{mtower}) the ADC looks like
\begin{equation}
V(t)=c~t^{-m}\quad (m=n\alpha)\, .
\end{equation}
In terms of the canonical field basis this is identical to an exponential potential of the form
\begin{equation}
V(\phi)=c~\exp\left(-n \sqrt\alpha \phi\right)\, .
\end{equation}

Let us briefly discuss what kind of wave functions are solutions of
the Schr\"odinger equation (\ref{schrodinger}). Actually, this potential is known as the Liouville potential and the  Schr\"odinger equation with this potential is exactly solvable.
The solutions for the wave function are given in terms of modified Bessel functions. For $\phi\rightarrow\infty$, they behave like plane waves: the form discussed before.
For the AdS case, the  potential is attractive, 
whereas 
for the de Sitter case, the potential is repulsive. One relevant example is the dark dimension scenario~\cite{Montero:2022prj}
with $n=4$ and $c^{1/4}\simeq 10^{-1}-10^{-3}$; the size $R=1/m_t$ of the dark dimension is in the micron range.
For one dark dimension ($\alpha=3/2$), one then gets that 
\begin{equation}
V(t)=c~t^{-6}=c~\exp\bigl(-2\sqrt 6\phi\bigr)
\,,
\end{equation}
and for two dark dimensions ($\alpha=1$)~\cite{Anchordoqui:2025nmb}, the potential becomes
\begin{equation}
V(t)=c~\exp\bigl(-4 \phi\bigr)\,.
\end{equation}
Both potentials are repulsive and do not lead to bound states. It will be interesting to study the solutions of the Schr\"odinger equation for these kind of potentials in more detail.

\section{$\bm{{\cal N}=2}$ Calabi-Yau string vacua }
\label{cubicF}

In this section, we look at a specific class of ${\cal N}=2$ models and also present their relation to the CRV pattern. 

\subsection{Type IIA Calabi-Yau compactifications and cubic prepotential}
\label{app:A}

Type IIA compactifications on CY manifolds in the large
K\"ahler moduli regime can be described by $\mathcal{N}=2$
supergravity, with a classical 
prepotential (or $F_0$ term) that is a cubic polynomial in the vector multiplet scalars,
\begin{equation}
    F_0(X^\Lambda)=-\frac16 C_{ijk} \frac{X^i X^j X^k}{X^0} \,,
\end{equation}
where $X^\Lambda = (X^0, X^i)$, the indices $i, j, k$ run over $1, ..,
h^{1,1} \equiv n_V$, and where $C_{ijk}$ are the triple intersection
numbers of the CY manifold. $F_0$ determines the classical
two-derivative action, which includes the Einstein-Hilbert term and
the kinetic energy of the scalar fields at quadratic level. The higher derivative corrections to the classical prepotential are
organized in a series expansion,
\begin{equation}
  F = F_0 + \sum_{g=1}^\infty F_g \,,
\end{equation}
where $F_g$ represents the contribution at genus $g$, corresponding to
$g$-loop effects in the supergravity theory. Each $F_g$ term
contributes to the effective action by generating higher-derivative
corrections, beyond the two-derivative level. 

Defining ${\cal F}_0(z)=F_0(X)/(X^0)^2$, we can write the classical prepotential as
\begin{equation}
    {\cal F}_0(z)= -\frac16 C_{ijk} z^i z^j z^k\, ,\label{calprep}
\end{equation}
where the physical scalars of the vector multiplets can be parametrized as
\begin{equation}
    z^i = \frac{X^i}{X^0} \, .
\end{equation}
We recall that $(X^\Lambda, F_{0,\Lambda})$ are symplectic sections, with 
\begin{align}
    F_{0,\Lambda}= \frac{\partial F_0}{\partial X^\Lambda} =
  X^0\left(\frac16 C_{ijk}z^i z^j z^k,-\frac12 C_{ijk} z^jz^k\right)
  \, .
\end{align}
Hyper multiplets are not described by the prepotential. 

The K\"ahler potential is given by
\begin{align}
    K &= -\log\left[i\left(\bar X^\Lambda F_{0,\Lambda} - X^\Lambda
        \bar F_{0,\Lambda}\right)\right] \nonumber \\
    &=-\log\left[\frac i6 C_{ijk}(z^i-\bar z^i)(z^j-\bar z^j)(z^k-\bar
      z^k) X^0\bar X^0\right] \nonumber \\
    &=-\log\left[\frac43 C_{ijk}t^it^jt^k \, X^0\bar X^0\right]
      \nonumber \\
    &=-\log\left[8\mathcal{V}X^0 \bar X^0\right],
  \end{align}
where we split the complex scalars into real and imaginary parts, 
\begin{equation}
    z^i = b^i + i t^i,
\end{equation}
and 
\begin{equation}
\mathcal{V} = \frac16 \int_{CY} J \wedge J \wedge J = \frac16 C_{ijk}t^it^jt^k
\end{equation}
is the overall classical volume of the
CY with K\"ahler 2-form $J$.
The associate KK mass scale, related to the overall 6-dimensional
volume is given by 
\begin{equation}
m_{KK}= \mathcal{V}^{-1/6}\, .
\end{equation}
Later, we will consider the case of anisotropic internal CY manifolds and their associated KK tower mass scales $m_t$ in more detail, as well the so-called emergent string limit.

But before that,  one can calculate
\begin{align}
K_i &=\partial_iK= -\frac i2 e^K C_{ijk}    (z^j-\bar z^j)(z^k-\bar
      z^k) \, X^0\bar X^0 \nonumber \\ &= 2i e^KC_{ijk} t^j t^k \, X^0\bar X^0,\\
K_{\bar\imath} &=\partial_{\bar\imath}K= \frac i2 e^K C_{ijk}
                 (z^j-\bar z^j)(z^k-\bar z^k) \, X^0\bar X^0 \nonumber
  \\ &= -2i  e^K C_{ijk} t^j t^k\, X^0\bar X^0,
\end{align}
and then the K\"ahler metric is explicitly
\begin{align}
g_{i\bar \jmath} &= K_i  K_{\bar \jmath} + i e^K C_{ijk} (z^k-\bar
                   z^k)\, X^0\bar X^0 \, .
\end{align}
Asking for the volumes of all complex submanifolds of the CY to be positive defines a region of the moduli space known as K\"ahler cone. For the supergravity approximation to be reliable, one should remain within the bulk of this region.
It is convenient to pass to the real basis of fields $t^i$, since
their axionic partners $b^i$ do not enter the K\"ahler
potential. Hence, throughout we set $b^i=0$.

\subsection{Three infinite distance limits and their associated  mass scales}

While remaining in the large volume regime, infinite distance limits in the vector multiplet moduli space of these theories correspond to sending a subset of the saxions to infinity, i.e.~$t^i  \rightarrow \infty$ with $i=1,\ldots,\mathbf{n}$ and $\mathbf{n}\leq h_{1,1}$, while all the axions are kept fixed. In general, the other saxions can be kept fixed at very large values--so as to stay in the large volume regime--or sent to infinity at slower rates. 
We will further specify the prepotential and we will consider three particular infinite distance limits of going to the asymptotic boundary of the moduli space.
Concretely, we will consider the simple case of three K\"ahler moduli
$z^i$ ($i=1,2,3$) and a classical prepotential with non-vanishing intersection number $C_{123}=1$ of the form
\begin{equation}
F_0(X)= -  z^1z^2z^3 (X^0)^2\, .
\end{equation}

The infinite-distance limits have been carefully classified in~\cite{Grimm:2018ohb,Corvilain:2018lgw,Lee:2019wij} and in the context
of the CRV pattern in~\cite{Castellano:2023stg,Castellano:2023jjt,Etheredge:2024tok,Etheredge:2025ahf}.
In addition, the relation to classical black holes as probes of three infinite distance limits was discussed in~\cite{Calderon-Infante:2025pls}.
Following the notation in \cite{Grimm:2018ohb}, we will denote them as Type IV, III and II. 
Let us summarize the three asymptotic limits in turn.

\vskip0.3cm

\paragraph{Type IV: Decompactification to five dimensions} \mbox{ } \\
\vskip0.3cm
Geometrically, this type of limit arises when the CY volume blows up uniformly. That is by identifying all three moduli, and we have $t^i \sim t\to \infty$ such that
\begin{equation}
    \mathcal V \sim t^3 \, .
\end{equation}
In this limit, the prepotential simply becomes
\begin{equation}
{\cal F}_0(t)= it^3\, .\label{f01}
\end{equation}

Furthermore, the mass scale of the related  tower in 4D Planck units is now given as
\begin{equation}
    \frac{(m_{t})_{\rm IV}}{M_{p}}  \sim t^{-3/2} \, .
    \label{kk-scale-iv-limit}
\end{equation}
Comparing with the prepotential we see that 
\begin{equation}
    (m_{t})_{\rm IV}  \sim (\Im{\cal F}_0)^{-1/2} \,.
    \label{kk-scale-iv-limit1}
\end{equation}

\vskip0.3cm

\paragraph{Type III: Decompactification to six dimensions} \mbox{ } \\
\vskip0.3cm
Type III limits are allowed when the CY admits an elliptic fibration. In particular, the infinite distance limit corresponds to blowing up the four-dimensional base while the volume of the elliptic fiber remains constant in string units. Hence, the CY volume satisfies
\begin{equation}
    \mathcal V \sim t^2 \, .
\end{equation}
This can be achieved by setting one modulus to a constant, e.g. $t_1=1$, and identifying the remaining two: $t_2=t_3=t$.
Then the prepotential in \eqref{calprep} is \emph{quadratic} in the moduli whose saxions are being sent to infinity:
\begin{equation}
{\cal F}_0(t)=it^2\, . \label{f02}
\end{equation}

The leading tower is composed by D0-branes and D2-branes wrapping the elliptic fiber, such that this limit corresponds to a decompactification to the 6D theory obtained by compactifying F-theory on the same CY. The mass scale of this KK tower is given by
\begin{equation}
    \frac{(m_{t})_{\rm III}}{M_{p}} \sim  t^{-1}\, .
    \label{kk-scale-iii-limit}
\end{equation}
Comparing with the prepotential we again see that 
\begin{equation}
    (m_{t})_{\rm III}  \sim (\Im{\cal F}_0)^{-1/2} \,.
    \label{kk-scale-iv-limit2}
\end{equation}

\vskip0.3cm

\paragraph{Type II: Emergent string limit} \mbox{ } \\
\vskip0.3cm
Finally, Type II limits can happen when the CY admits a K3 or an Abelian (i.e. ${\mathbb T}^4$) fibration. In this case, the infinite distance limit is reached when the volume of the two-dimensional base blows up while that of the fibration remains fixed in string units. That is,
\begin{equation}
    \mathcal V \sim t \, .
\end{equation}
Something similar holds for the prepotential in \eqref{calprep}, where we now fix $t_1=t_2=1$, and sent $t_3=t\rightarrow\infty$. Then the prepotential
 is found to be \emph{linear} in the modulus whose saxion is being sent to infinity:
 \begin{equation}
{\cal F}_0(t)= it\, .\label{f03}
\end{equation}
 In this limit, the lightest object is given by an NS5-brane wrapping the fiber.
  The tower scale satisfies
\begin{equation} \label{eq:scale-TypeII}
    \frac{(m_t)_{\rm II} }{M_{p}} \sim  t^{-1/2} \, .
\end{equation}
One more time we recover the relation  
\begin{equation}
    (m_t)_{\rm II}  \sim (\Im{\cal F}_0)^{-1/2} \,.
    \label{kk-scale-iv-limit3}
\end{equation}

\medskip

In summary, as one can observe, the different types of limits are nicely distinguished from the EFT perspective by having a cubic, quadratic or linear behavior of the prepotential with the moduli whose saxions are being sent to infinity. 
On the other hand, 
we like to highlight that the associated tower mass scale $m_t$ enjoys a universal relation to the prepotential in any of the three limits of the form
\begin{equation}
   m_t (t) \sim (\Im{\cal F}_0(t))^{-1/2} \,.
    \label{kk-scale-iv-limitsuniversal}
\end{equation}

\vskip0.3cm

The species scale shows up suppressing the curvature-squared corrections with respect to the Einstein-Hilbert term \cite{vandeHeisteeg:2022btw, vandeHeisteeg:2023ubh, vandeHeisteeg:2023dlw}.
In four-dimensional ${\cal N}=2$ supergravity, the Wilson coefficient of this higher-curvature correction is given by the genus-one topological string free energy, $F_1(X)={\cal F}_1(z)$, such that
\begin{equation}
\Lambda_s^{-2}=N_s\sim {\cal F}_1(t^i)\, .
\end{equation}
For any asymptotic limit, ${\cal F}_1$ behaves linearly with $t^i$, and is given in terms of the second Chern number of the CY manifold as
\begin{equation}
N_s\sim \Im{\cal F}_1(t^i)\sim c_{2,i}t^i\, .\label{f1}
\end{equation}
With respect to the large field limit of the leading field (omitting the index $i$) we see that there is a universal behavior of the species number like
\begin{equation}\label{eq:species}
   N_s \sim c_2t \, ,
\end{equation}
regardless of the type of limit that is being explored. 

Using this result, we can now express the species scale in terms of the tower mass scale, first for the type IV limit as
\begin{equation}\label{species5d}
    \Lambda_s \sim \, (m_{t})_{\rm IV}^{1/3} M_{p}^{2/3} \, .
\end{equation}
Consequently, the species scale is given by the 5D Planck scale, i.e. $\Lambda_s \sim M_{p,5}$. 

Similarly, for the case III limit, we obtain
\begin{equation}
    \Lambda_s  \sim (m_{t})_{\rm III}^{1/2} M_{p}^{1/2} \, ,
\end{equation}
implying that the species scale corresponds to the 6D Planck scale,  and thus is determined by $\Lambda_s \sim M_{p,6}$.

Lastly, for the type II limit, we obtain that
\begin{equation}
    \Lambda_s  \sim (m_{t})_{\rm II}  \, ,
\end{equation}
confirming that this limit corresponds to an emergent string limit with $\Lambda_s\sim M_{\rm string}$.

\subsection{The CRV pattern and commutators in terms of Calabi-Yau periods}

We have now prepared the ground to express the CRV relation, as given in (\ref{CRVpattern}), in terms of the relevant CY quantities.
Recall that for the three different asymptotic limits IV, III and II, we found the following two universal relations
\begin{eqnarray}
   m_t (t) &\sim& (\Im{\cal F}_0(t))^{-1/2}\sim t^{-\alpha}\, ,\nonumber\\
  N_s(t)&\sim& \Im{\cal F}_1(t)\sim t\, .\label{towerspeciesno}
\end{eqnarray}
Then, we can rewrite
the left hand side of (\ref{CRVpattern}) 
as
\begin{eqnarray}
\frac{\nabla m_t}{m_t} \cdot \frac{\nabla N_s}{N_s}
&=& -{1\over 2}\frac{{\nabla \cal F}_{0}(t)}{{\cal F}_0(t)} \cdot \frac{\nabla t}{t} \nonumber\\
&=&-{1\over 2}\frac{\partial t^{\alpha}}{t^{\alpha}}\frac{\partial t}{t}g^{t t}\,.
\end{eqnarray}
Using the inverse K\"ahler metrics $g^{t t}=t^2/\alpha$
one easily confirms 
the CRV relation  for all three limits IV, III and II.

Let us now connect the Calabi-Yau periods to the basic commutation
relations of the previous section. Duplicating our course of action, we define for a particular tower with label $\alpha$ the following pair of conjugate variables in terms of the periods $(t,\tilde {\cal F}_{0,t})$:
\begin{eqnarray}
 N_s\simeq ~t 
\end{eqnarray}
and
\begin{eqnarray}
\tilde m_t\simeq ~\Im\tilde {\cal F}_{0,t}&=&{1\over 2}t^{-1-2\alpha}{d\Im{\cal F}_0\over d\tau}\nonumber\\
&=&{1\over 2}t^{-1-2\alpha}\Im{\cal F}_{0,t}\dot t\, .
\end{eqnarray}
Then one can show that indeed
\begin{equation}
\lbrack t,\Im\tilde {\cal F}_{0,t}\rbrack=i\, .
\end{equation}

We can try to generalize this for several towers with labels
$\alpha_i$ to obtain the following canonical commutation relations for
$(t^i,\tilde \Im{\cal F}_{0,i})$:
\begin{equation}
\lbrack t^i,\Im\tilde {\cal F}_{0,j}\rbrack=i\delta^i_j\, .\label{CYcom2}
\end{equation}
In terms of the periods $(\Im X^\Lambda,\Re F_{0,\Lambda})$ one has the following quantization condition
\begin{equation}
\lbrack \Im X^\Lambda,\Re\tilde {F}_{0,{\Lambda{'}}}\rbrack=i\delta^\Lambda_{\Lambda'} \, .\label{CYcom}
\end{equation}
Actually note that in this equation, the indices span over the following range: $\Lambda,\Lambda'=0,\dots ,n_V$.
The corresponding periods $(X^\Lambda,\tilde F_{0,\Lambda})$ contain one non-propagating scalar field, which we conveniently can set to one, e.g. $X^0=1$. It follows that its conjugate momentum is zero: $\tilde F_{0,0}\simeq \dot X^0=0$.
So with this choice the commutators in eq.~\eqref{CYcom} reduce to the ones in eq.~\eqref{CYcom2}.
One can also see that 
T- and S-duality transformations act on the periods $(X^\Lambda,\tilde F_{0,\Lambda})$ as $\mathrm{Sp}(2n_V+2)$ symplectic transformations~\cite{Ceresole:1995jg,deWit:1995dmj,Antoniadis:1995ct}.
These symplectic transformation also act in natural way on the $2n_V+2$ electric and magnetic field strengths, which include also the graviphoton and its magnetic dual.

\vskip0.5cm

Finally, concerning possible corrections to the number of species
$N_s$ and the tower mass scale $m_t$ inside the ${\cal N}=2$ moduli
space we now like to suggest that inside the moduli space, the tower
mass scale $m_t$ and the species number $N_s$ are still given as in
 \eqref{towerspeciesno}, however, with ${\cal F}_0$ and ${\cal F}_1$
being the full non-perturbative expressions, which contain all
exponentially suppressed instanton corrections. Hence we propose for the
general CRV pattern including instanton corrections the following relation:
\begin{equation}
\frac{\nabla m_t}{m_t} \cdot \frac{\nabla N_s}{N_s} = -{1\over 2}\frac{\nabla {\cal F}_0(t)}{{\cal F}_0(t)} \cdot \frac{\nabla {\cal F}_1(t)}{{\cal F}_1(t)} \, .\label{CRVperiods}
\end{equation}
It is tempting to speculate about a possible connection or generalization of this relation to the BCOV holomorphic anomaly equations \cite{Bershadsky:1993cx, Bershadsky:1993ta}.

\subsection{Connection to the OVV quantization}

At the end of this section, we want to make contact with the quantization
rules set up in OVV in~\cite{Ooguri:2005vr}.
In this paper it is argued that in the context of black hole quantization  the topological string partition function can be interpreted as a ``wave-function of the universe in the mini-superspace sector of physical superstring theory.'' 
Furthermore, a corresponding black hole entropy can be viewed as the square of this topological string wave-function.

Actually, in ~\cite{Ooguri:2005vr} a Dirac bracket among the CY periods of the following form arises [see eq.~(5.11)\footnote{Note that in our paper we are using a different convention compared to OVV, namely the real and imaginary parts of
the periods are swapped.}]:
\begin{equation}
\lbrack \Im X^{\Lambda},\Im F_{0,\Lambda'}\rbrack_{\rm Dirac} \sim i \delta_{\Lambda'}^{\Lambda}\,.
\end{equation}
Clearly, this OVV commutator exhibits a very similar structure as the commutator in \eqref{CYcom}, which we have derived from the canonical commutation procedure.

Note that, unlike \eqref{CYcom} and the other commutator relations discussed above, the OVV commutators do not involve time derivatives of the fields.
This is because they are restricted to BPS configurations that live on a reduced phase space constrained by the BPS equations.

We can also briefly mention the mini-superspace wave function
$\Psi_{\rm OVV}$ of OVV, which has the following schematic form
\begin{equation}
|\Psi_{\rm OVV}(Q,P)|^2 \sim \exp\bigl({\cal S}_{\rm BH}(Q,P)\bigr)\, .
\end{equation}
Here ${\cal S}_{\rm BH}=F_{\rm top}+\bar F_{\rm top}+\cdots$ is the entropy of an ${\cal N}=2$ black hole that follows the attractor flow.
Furthermore $F_{\rm top}=\sum_{g=0}^\infty F_g$ is the full topological string partition function.
We see that this wave function is quite similar to the species wave function \eqref{specieswave},
which we have discussed before.

\section{Summary and Conclusions}

We have taken the first steps for laying a foundation of {\sl Species
  Quantum Mechanics}. Our main arguments can be summarized as follows.

\vskip0.3cm
First we defined canonical conjugate pairs of operators in the following way:
\vskip0.4cm

\fbox{\parbox{7.5cm}{\sl Consider a tower of species that satisfies the swampland distance conjecture or respectively the emergent string conjecture and also the CRV 
pattern relation. Then (asymptotically), this tower equivalently defines a pair of conjugate variables $N_s$ and $\tilde m_t$ with $\lbrack N_s,\tilde m_t\rbrack=i$.}
}

\vskip0.4cm

This leads to an uncertainty relation basically saying that a large
species uncertainty implies a small uncertainty in the time variation
of the tower mass and vice versa. As a consequence, we also demonstrated that the CRV pattern is closely related to this pair of conjugate variables.\footnote{As a side note, it is worth mentioning that the commutator in (\ref{fhcom1}) always  equates to ($i$ times) the CRV pattern,
because this relation is a proven property of quantum mechanical
operators, as demonstrated in (\ref{fhcom}). Now, assuming that the CRV pattern holds, the operators within the commutator may be treated as conjugates, but of course we
can explicitely use the pattern in the asymptotic regime. We have then
examined the pattern for three
possible types of light towers in infinite-distance limits, where the pattern has been shown to be satisfied.} Furthermore we also investigated the corresponding canonical commutation relations
of Calabi-Yau period vectors in  ${\cal N}=2$ string compactifications. Here we highlighted a universal relation of the species number $N_s$ and the tower mass scale $m_t$ in terms of the genus zero and the genus one prepotentials ${\cal F}_0$ and ${\cal F}_1$, which 
led us to a proposal for the CRV pattern including instanton corrections.

\vskip0.2cm

Second, we identified
a species wave function of the universe that determines the quantum evolution of number of species and the tower mass scale in the following way:
\vskip0.4cm

\fbox{\parbox{7.5cm}
{\sl The species wave function satisfies a Schr\"odinger equation  
${\cal H} \psi_s = E\psi_s$, 
and particularly if the potential vanishes everywhere is proportional to 
$\psi_s(t)\sim\exp \left[ ik\log N_s(t) \right]\sim \exp(ik\log{\cal
  S}_{\rm BH})$.}
}

\vskip0.4cm
This wave function corresponds to a kind of mini-superspace wave
function of quantum gravity. It is important to stress
  that herein we do not consider the quantum gravity over the
  four-dimensional space-time,  but of the scalars that come from the
  compact part of the metric. It would be interested to see, which expectation values for $N_s$ and $m_t$ have high or low probability (see also~\cite{LopesCardoso:2006efn,LopesCardoso:2006xue}).
Furthermore, we noted that the species wave function $\psi_s$ is very
similar to the OVV wave function in  ${\cal N}=2$ string
compactifications. We also discussed some cases with non-vanishing
potentials like those characterizing the dark dimension scenario.

\vskip0.2cm

Third we  defined duality transformations in the following way:
\vskip0.4cm

\fbox{\parbox{7.5cm}
{\sl 
Symplectic transformations on the pairs of conjugate variables $(N_s,\tilde m_t)$ lead  to T-duality or S-duality transformations in the moduli space.}
}

\vskip0.4cm
These duality transformations also correspond to symplectic transformations on Calabi-Yau periods in ${\cal N}=2$ string compactifications.

\vskip0.3cm

Ultimately, the reader should be aware of a few caveats and questions
in our arguments: {\it (i)} In order to define the species quantum
mechanics, we simplified the discussion by
considering only one dimension.
It should be possible to extend the discussion to a more field theoretic framework in higher dimensions.
{\it (ii)}~There are corrections to the moduli dependence of the species tower  as well as to the CRV pattern in the interior of the moduli space. 
So, how do these corrections change the commutator relations? {\it
  (iii)}~We would like to have  a better understanding about the
corresponding species Hilbert space and the species wave functions. It
would be interesting  to use the species wave function to compute the
probability density in moduli space. For example, does this
probability density adhere to the (macroscopic) second law of species
thermodynamics, being high at the
boundary of the moduli space and low inside the moduli space?

Finally, it would be also interesting to consider other swampland
conjectures and bounds, like the $V'/V$-bound~\cite{Obied:2018sgi} of the de Sitter
conjecture~\cite{Obied:2018sgi,Dvali:2014gua, Dvali:2017eba,Dvali:2018jhn, Garg:2018reu,Ooguri:2018wrx,Dvali:2018fqu}.
These and other questions would be worth to pursue in the future.



\section*{Acknowledgements}

We very much thank Carmine Montella for many enlightening
discussions. We also thank Ivano Basile for valuable comments on the manuscript. The work of L.A.A. is supported by the U.S. National Science
Foundation (NSF Grant PHY-2412679). The work of D.L. is supported by
the Origins Excellence Cluster and by the German-Israel-Project (DIP)
on Holography and the Swampland.


\end{document}